\documentclass[12pt]{article}

\usepackage{amsfonts, amsmath}

\newcommand{\be}{\begin{equation}} \newcommand{\ee}{\end{equation}}

\begin{document}
\title{ Deformed density matrix, Density of
entropy and Information problem }
\thispagestyle{empty}

\author{A.E.Shalyt-Margolin\hspace{1.5mm}\thanks
{Fax +375 172 326075; e-mail:a.shalyt@mail.ru;alexm@hep.by}}
\date{}
\maketitle
 \vspace{-25pt}
{\footnotesize\noindent  National Center of Particles and High
Energy Physics, Bogdanovich Str. 153, Minsk 220040, Belarus\\
{\ttfamily{\footnotesize
\\ PACS: 03.65; 05.70
\\
\noindent Keywords: deformed density matrices, density of
entropy, information problem}}

\rm\normalsize \vspace{0.5cm}
\begin{abstract}
Quantum Mechanics at Planck scale is considered as a deformation
of the conventional Quantum Mechanics. Similar to the earlier works
of the author, the main object of deformation is the density matrix.
On this basis a notion of the entropy density is introduced that is
a matrix value used for a detail study of the Information Problem at
the Universe, and in particular, for the Information Paradox Problem.
\end{abstract}
\newpage
\section{Introduction}

This paper presents one of the concepts of Quantum Mechanics with
Fundamental Length (QMFL) considered as Quantum Mechanics at
Planck scale \cite{r1},\cite{r2} due to the Generalized
Uncertainty Relations (GUR) \cite{r4}-\cite{r12}, developed in
more early papers of the author with colleagues
\cite{r1},\cite{r2} as a concept of the entropy density.
Similarly, the main object is the deformed density matrix at
Planck scale (density pro-matrix).It is shown that indeed the
density of entropy is a matrix value:
\\
$$S^{\alpha_{1}}_{\alpha_{2}}=-Sp[\rho(\alpha_{1})\ln(\rho(\alpha_{2}))]=
-<\ln(\rho(\alpha_{2}))>_{\alpha_{1}}.$$
\\
The exact determination of matrix
$S^{\alpha_{1}}_{\alpha_{2}}$ is given in  section 3 of the paper.
Based on this approach, the value
\\$S_{\alpha}=-Sp[\rho(\alpha)\ln(\rho(\alpha))]=
-<\ln(\rho(\alpha))>_{\alpha}$ originally considered in
\cite{r1},\cite{r2} is nothing but a diagonal matrix element of
matrix  $S^{\alpha_{1}}_{\alpha_{2}}$. In section 3 it is shown
that this matrix is practically a matrix of the entropy densities
on the unit minimum area for different observers. Then it is used
for a detailed study of the Information Problem at the Universe,
and in particular for the Information Paradox Problem \cite{r3}.
This problem is reduced to comparison of the initial and final
densities of entropy for one and the same observer. It's shown
that according to the natural standpoint, there is no information
loss at the closed Universe. Similar to our previous works, we use
here the abbreviation QMFL \cite{r1},\cite{r2} for Quantum
Mechanics with Fundamental Length \cite{r4}-\cite{r12} and QM for
the conventional Quantum Mechanics \cite{r13}. In section 2 the
formalism of the density pro-matrix in QMFL is described in brief.
In conclusion a short analysis of the prospects for the proposed
approach is given with comparison to other methods.

\section{Deformed   Density Matrix in QMFL}

In this section the principal features of QMFL construction using
the deformed density matrix are briefly outlined  \cite{r2}. In
the notation system used for $\alpha = l_{min}^{2 }/x^{2 }$ $x$ is
the scale for the fundamental deformation parameter.

\noindent {\bf Definition 1.} (Quantum Mechanics with Fundamental
Length)

\noindent Any system in QMFL is described by a density pro-matrix
of the form $\rho(\alpha)=\sum_{i}\omega_{i}(\alpha)|i><i|$, where
\begin{enumerate}
\item $0<\alpha\leq1/4$.
\item The vectors $|i>$ form a full orthonormal system.
\item $\omega_{i}(\alpha)\geq 0$, and for all $i$  the
finite limit $\lim\limits_{\alpha\rightarrow
0}\omega_{i}(\alpha)=\omega_{i}$ exists.
\item
$Sp[\rho(\alpha)]=\sum_{i}\omega_{i}(\alpha)<1$,
$\sum_{i}\omega_{i}=1$.
\item For every operator $B$ and any $\alpha$ there is a
mean operator $B$ depending on  $\alpha$:\\
$$<B>_{\alpha}=\sum_{i}\omega_{i}(\alpha)<i|B|i>.$$
\end{enumerate}
Finally, in order that our definition 1 be in agreement
with the result of
(\cite{r2} section 2),
the following condition must be fulfilled:
\begin{equation}\label{U1}
Sp[\rho(\alpha)]-Sp^{2}[\rho(\alpha)]\approx\alpha.
\end{equation}
Hence we can find the value for $Sp[\rho(\alpha)]$ satisfying the
condition of definition 1:
\begin{equation}\label{U2}
Sp[\rho(\alpha)]\approx\frac{1}{2}+\sqrt{\frac{1}{4}-\alpha}.
\end{equation}

According to point 5, $<1>_{\alpha}=Sp[\rho(\alpha)]$. Therefore
for any scalar quantity $f$ we have $<f>_{\alpha}=f
Sp[\rho(\alpha)]$. In particular, the mean value
$<[x_{\mu},p_{\nu}]>_{\alpha}$ is equal to
\\
$$<[x_{\mu},p_{\nu}]>_{\alpha}= i\hbar\delta_{\mu,\nu}
Sp[\rho(\alpha)]$$
\\
We denote the limit $\lim\limits_{\alpha\rightarrow
0}\rho(\alpha)=\rho$ as the density matrix. Evidently, in the
limit $\alpha\rightarrow 0$ we return to QM.

\renewcommand{\theenumi}{\Roman{enumi}}
\renewcommand{\labelenumi}{\theenumi.}
\renewcommand{\labelenumii}{\theenumii.}

It should be noted that:

\begin{enumerate}
\item The above limit covers both Quantum
and Classical Mechanics. Indeed, since $\alpha\sim L_{p}^{2 }/x^{2
}=G \hbar/c^3 x^{2}$, we obtain:
\begin{enumerate}
\item $(\hbar \neq 0,x\rightarrow
\infty)\Rightarrow(\alpha\rightarrow 0)$ for QM;
\item $(\hbar\rightarrow 0,x\rightarrow
\infty)\Rightarrow(\alpha\rightarrow 0)$ for Classical Mechanics;
\end{enumerate}
\item As a matter of fact, the deformation parameter $\alpha$
should assume the value $0<\alpha\leq1$.  As seen from
(\ref{U2}), however, $Sp[\rho(\alpha)]$ is well defined only for
$0<\alpha\leq1/4$. That is if $x=il_{min}$ and $i\geq 2$, then
there is no any problem.  At the point of $x=l_{min}$ there is
a singularity related to the complex values following from
$Sp[\rho(\alpha)]$ , i.e. to the impossibility of obtaining a
diagonalized density pro-matrix at this point over the field of
real numbers. For this reason definition 1 has no sense at the
point $x=l_{min}$.
\item We consider possible solutions for (\ref{U1}).
For instance, one of the solutions of (\ref{U1}), at least to the
first order in $\alpha$, is $$\rho^{*}(\alpha)=\sum_{i}\alpha_{i}
exp(-\alpha)|i><i|,$$ where all $\alpha_{i}>0$ are independent of
$\alpha$ and their sum is equal to 1. In this way
$Sp[\rho^{*}(\alpha)]=exp(-\alpha)$. We can easily verify that
\begin{equation}\label{U3}
Sp[\rho^{*}(\alpha)]-Sp^{2}[\rho^{*}(\alpha)]=\alpha+O(\alpha^{2}).
\end{equation}
 Note that in the momentum representation $\alpha\sim p^{2}/p^{2}_{pl}$,
where $p_{pl}$ is the Planck momentum. When present in the matrix
elements, $exp(-\alpha)$ can damp the contribution of great
momenta in a perturbation theory.
\end{enumerate}

\section{ Entropy Density Matrix and Information Loss Problem }

In \cite{r2} the authors were too careful, when introducing for density
pro-matrix $\rho(\alpha)$ the value $S_{\alpha}$ generalizing the
ordinary statistical entropy:
\\
 $$S_{\alpha}=-Sp[\rho(\alpha)\ln(\rho(\alpha))]=
 -<\ln(\rho(\alpha))>_{\alpha}.$$
\\
In \cite{r1},\cite{r2} it was noted that $S_{\alpha}$ means
of the entropy density   on a unit  minimum area depending on
the scale. In fact a more general concept
accepts the form of the entropy density matrix:
\begin{equation}\label{U4}
S^{\alpha_{1}}_{\alpha_{2}}=-Sp[\rho(\alpha_{1})\ln(\rho(\alpha_{2}))]=
-<\ln(\rho(\alpha_{2}))>_{\alpha_{1}},
\end{equation}
where $0< \alpha_{1},\alpha_{2}\leq 1/4.$
\\ $S^{\alpha_{1}}_{\alpha_{2}}$ has a clear physical meaning:
the entropy density is computed  on the scale associated with
the deformation parameter $\alpha_{2}$ by the
observer who is at a scale corresponding to the deformation parameter
$\alpha_{1}$. Note that with this approach the diagonal element
$S_{\alpha}=S_{\alpha}^{\alpha}$,of the described matrix
$S^{\alpha_{1}}_{\alpha_{2}}$ is the density of
entropy measured by the observer  who is at the same scale  as the
measured object associated with the deformation parameter
$\alpha$. In \cite{r2} section 6 such a construction
was used implicitly in derivation of the semiclassical
Bekenstein-Hawking formula for the Black Hole entropy:

a) For the initial (approximately pure) state
\\
$$S_{in}=S_{0}^{0}=0$$
\\
b) Using the exponential ansatz(\ref{U3}),we obtain:
\\
$$S_{out}=S^{0}_{\frac{1}{4}}=-<ln[exp(-1/4)]\rho_{pure}>=-<\ln(\rho(1/4))>
=\frac{1}{4}.$$
\\
So increase in the entropy density for an external observer
at the large-scale limit is 1/4. Note that increase of
the entropy density  (information loss) for the
observer that is crossing the horizon of the black hole's events
and moving with the information flow to singularity will be smaller:
\\
$$S_{out}=S_{\frac{1}{4}}^{\frac{1}{4}}=-Sp(exp(-1/4)
ln[exp(-1/4)]\rho_{pure})=-<\ln(\rho(1/4))>_{\frac{1}{4}} \approx
0.1947 .$$
It is clear that this fact may be interpreted as follows:
for the observer moving
together with information its loss can  occur only at
the transition to smaller scales, i.e. to greater
deformation parameter $\alpha$. \\
\\ Now we consider the general Information Problem.
Note that with the classical Quantum Mechanics (QM) the entropy
density matrix $S^{\alpha_{1}}_{\alpha_{2}}$ (\ref{U4}) is reduced
only to one element $S_{0}^{0}$ and so we can not test anything.
Moreover, in previous works relating the quantum mechanics of
black holes and information paradox \cite{r3},\cite{r18,r19} the
initial and final states when a particle hits the
 hole are treated proceeding from different theories(QM and QMFL respectively):
\\
\\
(Large-scale limit, QM,
 density matrix) $\rightarrow$ (Black Hole, singularity, QMFL,
density pro-matrix),
\\
\\
Of course in this case any conservation of information is
impossible as these theories are based on different concepts of entropy.
Simply saying, it is incorrect to compare the entropy interpretations
of two different theories (QM and QMFL,
where this notion is originally differently understood.
So the chain above must be symmetrized by accompaniment
of the arrow on the left ,so in an ordinary situation we have a
chain:
\\
\\
(Early Universe, origin singularity, QMFL, density pro-matrix)
$\rightarrow$
\\ (Large-scale limit, QM,
 density matrix)$\rightarrow$ (Black Hole, singularity, QMFL,
density pro-matrix),
\\
\\
So it's more correct to compare entropy close to the initial and final
(Black hole) singularities. In other words, it is necessary to
take into account not only the state, where information disappears,
but also that whence it appears. The question arises, whether the
information is lost in this case for every separate observer. For
the event under consideration this question sounds as follows: are the
entropy densities S(in) and S(out) equal for every separate observer?
It will be shown that in all conceivable cases they are equal.

1) For the observer in the large-scale limit (producing
measurements in the semiclassical approximation) $\alpha_{1}=0$
\\
\\
$S(in)=S^{0}_{\frac{1}{4}}$ (Origin singularity)
\\
\\
$S(out)=S^{0}_{\frac{1}{4}}$ (Singularity in Black Hole)
\\
\\
So $S(in)=S(out)=S^{0}_{\frac{1}{4}}$. Consequently, the initial
and final densities of entropy are equal and there is no any
information loss.
\\
2) For the observer moving together with the information flow in
the general situation  we have the chain:
\\
$$S(in)\rightarrow S(large-scale)\rightarrow S(out),$$
\\
where $S(large-scale)=S^{0}_{0}=S$. Here $S$ is the ordinary entropy
at quantum mechanics(QM), but
$S(in)=S(out)=S^{\frac{1}{4}}_{\frac{1}{4}}$,value considered in
QMFL. So in this case the initial and final densities of entropy are
equal without any loss of information.
\\
3) This case is a special case of 2), when we do not come out of
the early Universe considering the processes with the participation
of black mini-holes only. In this case the originally specified
chain becomes shorter by one section:
\\
\\
(Early Universe, origin singularity, QMFL, density
pro-matrix)$\rightarrow$ (Black Mini-Hole,
singularity, QMFL, density pro-matrix),
\\
\\
and member $S(large-scale)=S^{0}_{0}=S$ disappears at the
corresponding chain of the entropy density associated with
the large-scale consideration:
\\
$$S(in)\rightarrow S(out),$$
\\
It is, however, obvious that in case
$S(in)=S(out)=S^{\frac{1}{4}}_{\frac{1}{4}}$ the density of
entropy is preserved. Actually this event was mentioned in section
5 \cite{r2},where from the basic principles it has been found that
black mini-holes do not radiate, just in agreement with the
results of other authors \cite{r14}-\cite{r17}.
\\ As a result, it's possible to write briefly
\\
$$S(in)=S(out)=S^{\alpha}_{\frac{1}{4}},$$
\\
where $\alpha$ - any value in the interval $0<\alpha\leq 1/4.$
\\ It should be noted that in terms of deformation the Liouville's
equation (section 4 \cite{r2}) takes the form:
\\
$$\frac{d\rho}{dt}=\sum_{i}
\frac{d\omega_{i}[\alpha(t)]}{dt}|i(t)><i(t)|-i[H,\rho(\alpha)]=
\\d[ln\omega(\alpha)]\rho (\alpha)-i[H,\rho(\alpha)].$$
\\
The main result of this section is a necessity to
account for the member $d[ln\omega(\alpha)]\rho (\alpha)$,deforming
the right-side expression of $\alpha\approx 1/4$.

\section{Conclusion}

Note that the proposed approach enables a study of the Information
Problem at the Universe from the basic principles proceeding from
the existing two types of quantum mechanics only: QM that
describes nature at the well known scales and QMFL at Planck
scales \cite{r1},\cite{r2},\cite{r11}. The author is of the
opinion that further development of this approach will allow to
research the information problem in greater detail. Besides, it is
related to other methods, specifically to the holographic
principle \cite{r20}, as the entropy density matrix studied in
this work is related to the two-dimensional objects. This paper is
a slightly revised variant of the preprint \cite{r21}.


\end{document}